\documentclass[showpacs,amsmath,amssymb,twocolumn,pra,superscriptaddress]{revtex4-1}
\usepackage{amssymb}
\usepackage[dvips]{graphicx}
\usepackage{enumerate}
\usepackage{epsfig}
\usepackage{subfigure}
\usepackage{xcolor}
\usepackage[T1]{fontenc}
\usepackage{fullpage}
\usepackage{amsthm,amsfonts,amssymb,amscd,mathrsfs,xspace,framed}
\usepackage{amsmath}
\usepackage{color}
\usepackage{setspace}
\usepackage{url}
\usepackage{wrapfig}
\usepackage{tikz}
\usepackage{enumitem}
\usepackage{array}
\DeclareMathOperator{\sinc}{sinc}

\begin{document}

\title{Wave Function Engineering for Spectrally-Uncorrelated Biphotons in the Telecommunication Band based on a Machine-Learning Framework}

\author{Chaohan Cui}
\email{email: chaohancui@email.arizona.edu}
\affiliation{College of Optical Sciences, The University of Arizona, Tucson, Arizona 85721, USA}
\author{Reeshad Arian}
\affiliation{Department of Electrical and Computer Engineering, The University of Arizona, Tucson, Arizona 85721, USA}
\affiliation{Department of Mathematics and Computational Sciences, The University of Arizona, Tucson, Arizona 85721, USA}
\author{Saikat Guha}
\affiliation{College of Optical Sciences, The University of Arizona, Tucson, Arizona 85721, USA}
\affiliation{Department of Electrical and Computer Engineering, The University of Arizona, Tucson, Arizona 85721, USA}
\author{N. Peyghambarian}
\affiliation{College of Optical Sciences, The University of Arizona, Tucson, Arizona 85721, USA}
\affiliation{Department of Materials Science and Engineering, The University of Arizona, Tucson, Arizona 85721, USA}
\author{Quntao Zhuang}
\affiliation{Department of Electrical and Computer Engineering, The University of Arizona, Tucson, Arizona 85721, USA}
\affiliation{College of Optical Sciences, The University of Arizona, Tucson, Arizona 85721, USA}
\author{Zheshen Zhang}
\affiliation{Department of Materials Science and Engineering, The University of Arizona, Tucson, Arizona 85721, USA}
\affiliation{College of Optical Sciences, The University of Arizona, Tucson, Arizona 85721, USA}

\begin{abstract}
Indistinguishable single photons are key ingredient for a plethora of quantum information processing applications ranging from quantum communications to photonic quantum computing. A mainstream platform to produce indistinguishable single photons over a wide spectral range is based on biphoton generation through spontaneous parametric down-conversion (SPDC) in nonlinear crystals. The purity of the SPDC biphotons, however, is limited by their spectral correlations. Here, we present a design recipe, based on a machine-learning framework, for the engineering of biphoton joint spectrum amplitudes over a wide spectral range. By customizing the poling profile of the KTiOPO$_4$ (KTP) crystal, we show, numerically, that spectral purities of 99.22\%, 99.99\%, and 99.82\% can be achieved, respectively, in the 1310-nm, 1550-nm, and 1600-nm bands after applying a moderate 8-nm filter. The machine-learning framework thus enables the generation of near-indistinguishable single photons over the entire telecommunication band without resorting to KTP crystal's group-velocity-matching wavelength window near 1582 nm.
\end{abstract}

\maketitle
\section{Introduction}
Quantum information science is an emerging area of study that creates new opportunities for the next-generation communication, computing, and sensing applications. Photons are unique quantum-information carriers as they can be transmitted over long distances for entanglement distribution \cite{yin2017satellite,chou2007functional}, secure communication \cite{liao2017satellite,boaron2018secure}, and sensing \cite{giovannetti2011advances,zhang2015entanglement}. In addition, single photons would be critical resources in near-term quantum-computing devices for, e.g., Boson Sampling \cite{tichy2015sampling,neville2017classical,renema2018efficient,niu2018qudit,bell2018multiphoton}, to demonstrate a performance advantage over any classical computing platforms, a.k.a., the quantum supremacy.

Specifically, the quantum internet \cite{kimble2008quantum,wehner2018quantum} will be empowered by single photons that herald the creation of entanglement between network nodes at a distance \cite{duan2001long}. Such a capability underpins distributed quantum computing \cite{lim2005repeat,serafini2006distributed,kok2007linear} and distributed quantum sensing \cite{proctor2018multiparameter,ge2018distributed,zhuang2018distributed,xia2019repeater}. The quality of the heralded entanglement, produced by interfering two single photons on a beam splitter to erase the which-way information, is critically dependent on the indistinguishability of the two photons. To ensure high indistinguishability, each interfering photon needs to be in a pure state in the spectral, temporal, spatial, and polarization domains. In addition, it is desirable that the single photons situate in the telecommunication band to leverage the abundant modulation, transmission, and detection devices for long-distance quantum communications.

Nonlinear crystals are widely employed to produce entangled and heralded single photons \cite{fradkin1999tunable,grice2001eliminating,mosley2008heralded,branczyk2011engineered}. Compared to solid-state single-photon emitters such as quantum dots and nitrogen-vacancy centers, nonlinear crystals enjoy room-temperature operations, the capability of generating photons in the telecommunication band, and the absence of spectral diffusion that degrades the purity of the produced photons. KTiOPO$_4$ (KTP), in this regard, is a widely used nonlinear crystal material by virtue of its high nonlinearity and broad transparency window. In particular, KTP possesses a group-velocity-matching (GVM) wavelength around 1582 nm \cite{grice2001eliminating} vouchsafed by its material dispersion. Such a unique property has been harnessed to generate spectrally-uncorrelated biphotons near the telecommunication c-band at 1550 nm. To achieve phase matching, two crystal poling strategies have been pursued. In the conventional periodic-poling strategy, shown in Fig.~\ref{fig:Po} (top), the positive and negative polarities each constitutes half of the duty cycle in each poling period, resulting in a $\sinc$ phase-matching profile whose side lobes limit the spectral purity. As a result, a narrowband filter is typically employed to cut off the side lobes, at the cost of reducing the flux and the heralding efficiency. To mitigate the limitation of periodic poling, Dixon {\em et al.} introduced a customized aperiodic poling profile, illustrated in Fig.~\ref{fig:Po} (bottom), to achieve a Gaussian phase-matching profile at the GVM wavelength \cite{dixon2013spectral}. In conjunction with a Gaussian-spectrum pump, a 99.5\% spectral purity was measured after applying a 8.5-nm full width at half maximum Gaussian spectral filter. The spectral purity of the biphotons produced in Dixon {\em et al.}'s scheme, however, degrades to 97.12\% after applying a 40-nm filter in the 1550-nm band due to the deviation from the GVM wavelength. In follow-up works ~\cite{dosseva2016shaping, tambasco2016domain, graffitti2017pure,graffitti2018independent,graffitti2018design,ansari2018tailoring}, several poling-design optimization approaches were introduced to improve upon  Ref.~\cite{dixon2013spectral}'s spectral purity. Like Dixon {\em et al.}'s scheme, these approaches require operating in the vicinity of the GVM wavelength, which precludes them from designing crystals for spectrally-uncorrelated biphoton generation over the entire telecommunication window from $\sim$1300 nm to 1600 nm. In addition, these approaches rely on binary optimization that limits the achievable purity, due to a lack of access to the full parameter space. To generate high purity biphotons at wavelengths apart from 1582 nm, KTP's output spectral purity is limited to merely $\sim$81\% \cite{jin2013widely}, which is insufficient for many applications. Apart from KTP, References ~\cite{jin2016spectrally,laudenbach2017numerical,jin2019theoretical} complied a list of other nonlinear materials, each operating at a specific wavelength dictated by its GVM property. However, a scheme for generating high spectrally-uncorrelated biphotons at any target wavelength remains elusive.

Here, we present a general machine-learning framework that seeks the optimum poling design for the generation of spectrally-uncorrelated biphotons. Unlike prior works, our approach exploits an optimization fully empowered by machine learning to obviate the need for the GVM property of nonlinear crystals. This yields a purity in excess of 99.8\% with a 8-nm filter over 1310 nm to 1600 nm. In particular, a 99\% spectral purity in the 1550-nm band is achieved after applying a 40-nm wideband filter that nicely maintains the flux and the heralding efficiency. Our result demonstrates the power of machine learning in tackling hard quantum-information problems.

\section{Biphoton JSA and phase matching}
In spontaneous parametric down-conversion (SPDC), the generated biphotons can be represented in the frequency domain as a superposition of different frequency modes:
\begin{equation}
    |\psi\rangle_{SI} = \iint d\omega_S d\omega_I f(\omega_S,\omega_I) \hat{a}^\dag_{\omega_S}\hat{a}^\dag_{\omega_I} |0\rangle_S|0\rangle_I.
\end{equation}
Here, $\hat{a}^\dag_{\omega_S}$ is the creation operator for the signal photon, $\hat{a}^\dag_{\omega_I}$ is the creation operator for the idler photon, and $f(\omega_S,\omega_I)$ is the biphoton joint spectrum amplitude (JSA) that entails complete information about the spectral-temporal properties of the photon pair. The biphoton JSA is determined by the pump spectrum and the properties of the nonlinear crystal by $f(\omega_S,\omega_I)\propto \alpha(\omega_P)G(\Delta k)$, where $\omega_P=\omega_S+\omega_I$ relates the pump, signal, and idler frequencies by energy conservation, $\alpha(\omega_P)$ describes the pump spectral profile, and $G(\Delta k)$ encompasses information about the phase-matching properties of the nonlinear crystal \cite{grice2001eliminating}. Specifically, the phase-matching function
\begin{equation}
G(\Delta k)= \frac{1}{L}\int_0^L g(z) \exp(-i\Delta k z) dz,    
\end{equation}
where $g(z) = \{1,-1\}$ describes the poling profile along the propagation $z$ axis. The phase mismatch
\begin{equation}
\begin{aligned}
& \Delta k(\omega_S,\omega_I) =k_P(\omega_P)-k_S(\omega_S)-k_I(\omega_I)\\ &
= 2\pi n(\omega_P)/\lambda_P - 2\pi n(\omega_S)/\lambda_S - 2\pi n(\omega_I)/\lambda_I,
\end{aligned}
\end{equation}
where $n(\omega)$ is the frequency-dependent refractive index that determines the material dispersion, and $\lambda_P$, $\lambda_S$, and $\lambda_I$ are the pump, signal, and idler wavelengths \cite{grice2001eliminating,dixon2013spectral,graffitti2018design}. To quantitatively describe the spectral correlation between the signal and idler, the JSA is decomposed \cite{dixon2013spectral,ansari2018tailoring} to $f(\omega_S,\omega_I)=\sum_n \xi_n\beta_{S,n}(\omega_S)\beta_{I,n}(\omega_I)$. Upon this, the purity is defined as $\mathcal{P}=\sum_n \xi_n^4/(\sum_n \xi_n^2)^2$. If $f(\omega_S,\omega_I)=\beta_S(\omega_S)\beta_I(\omega_I)$, i.e., $\mathcal{P}=1$, the JSA then describes a product state of spectrally-uncorrelated two photons, which are particularly useful for Boson sampling, entanglement distribution, and photonic quantum information processing \cite{niu2018qudit,sun2017entanglement,mosley2008heralded}. 

\begin{figure}
\centering
\includegraphics[width=8cm]{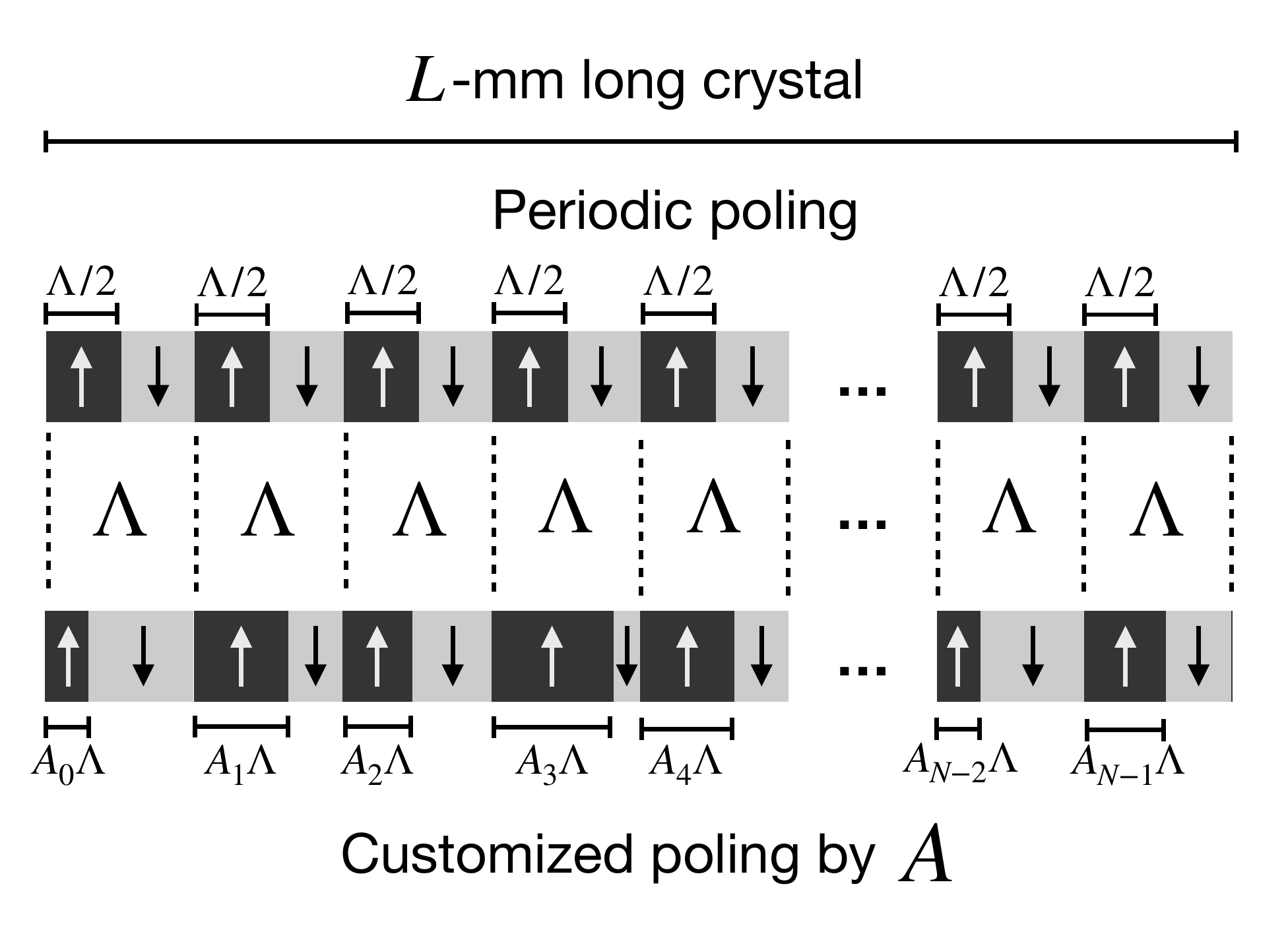}
\caption{\label{fig:Po} Poling profiles for a $L$-mm-long nonlinear crystal with $N$ poling periods: (top) periodic poling with a period of $\Lambda$; (bottom) customized poling with a profile embedded in the array $\boldsymbol{A}$.}
\end{figure}

To engineer a desired biphoton JSA $f(\omega_S,\omega_I)$, one has two tunable knobs: the pump spectral profile $\alpha(\omega_P)$ and the crystal poling profile $g(z)$. In this paper, we introduce a general recipe that harnesses a machine-learning framework to automate the design for $\alpha(\omega_P)$ and $g(z)$. Let us first formulate the JSA engineering problem and provide some insights into its connection with machine learning.

The poling profile embedded in $g(z)$ involves the poling period $\Lambda$ and an array $\boldsymbol{A}=\{A_i\}$ that specifies the duty cycle in each of the $N$ periods. $\Lambda$ is given by the phase-matching condition at pump's central frequency $\omega_{P_0}$ when signal and idler photons are wavelength degenerate at $\omega_{P_0}/2$: $\Lambda=2\pi/\Delta k_0$, where $\Delta k_0=\Delta k(\omega_S=\omega_I=\omega_{P_0}/2)$. $g(z)$ thus relates to $\Lambda$ and $\boldsymbol{A}$ by
\begin{equation}
\label{eq:gz}
g(z,\boldsymbol{A})=-1+2\sum_{j=0}^{N-1} \left[\Theta(z-j\Lambda)-\Theta(z-(j+A_j)\Lambda)\right]
\end{equation}
where $\Theta$ is the unit step function (see Fig.~\ref{fig:Po}). Plugging $g(z)$, given by Eq. \ref{eq:gz}, into the phase-matching function yields
\begin{equation}\begin{aligned}
G(\Delta k,\boldsymbol{A})&=\frac{1}{iL\Delta k}\sum_{j=0}^{N-1}[e^{-i\Lambda j\Delta k}\\
&+e^{-i\Lambda (j+1)\Delta k}-2e^{-i\Lambda (j+A_j)\Delta k}]
\end{aligned}\end{equation}

In practice, the design of $g(z)$ is limited by the minimum poling length $\Lambda_{\rm min}$. To accommodate the practical limitation, we modify the duty cycles as $A_i\in[\Lambda_{\rm min}/\Lambda, 1-\Lambda_{\rm min}/\Lambda]$ (the ratio of the positive polarity portion within each poling period) in tuning the JSA. 

\begin{figure}
\centering
\includegraphics[width=8cm]{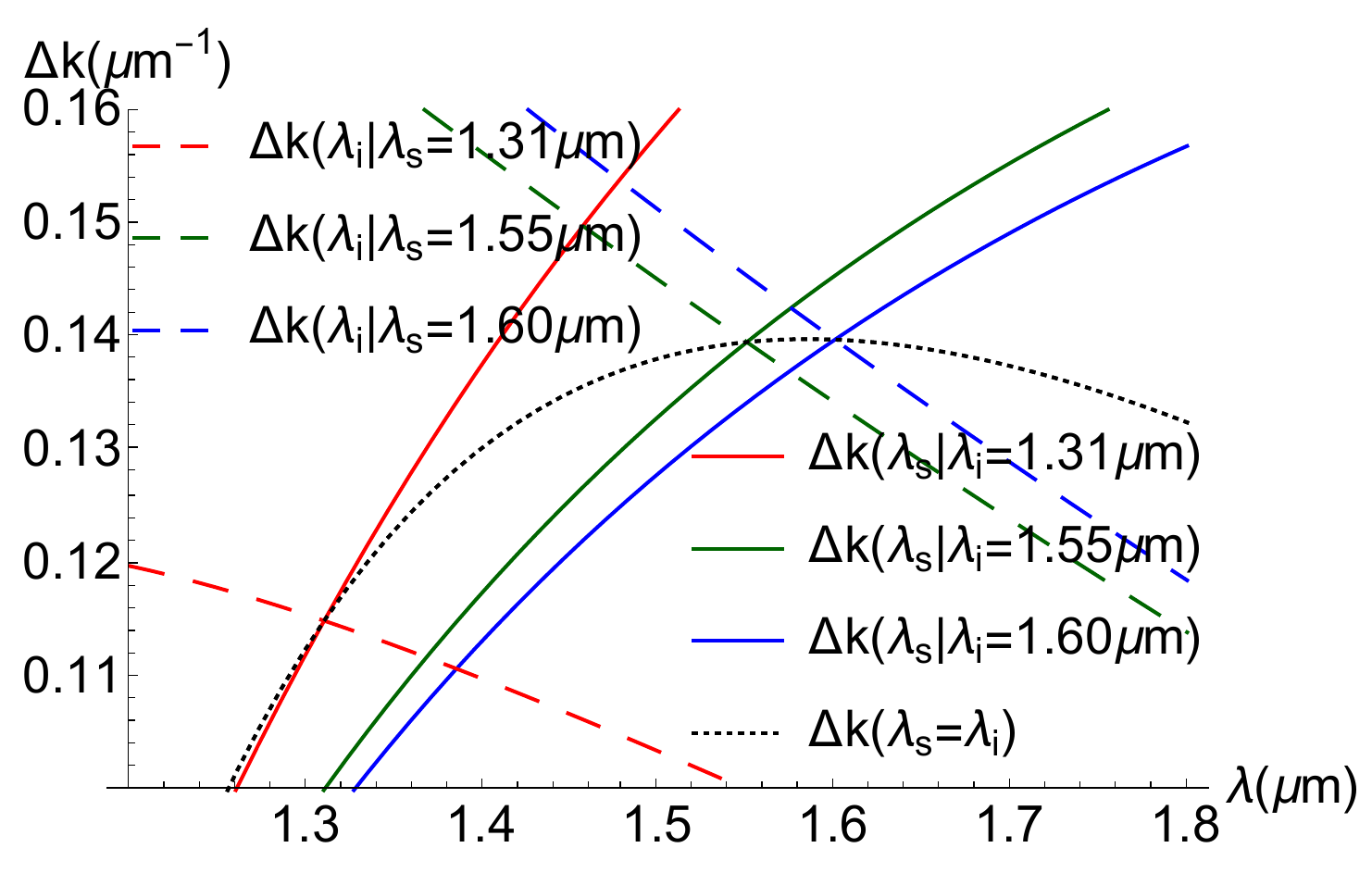}
\caption{\label{fig:DK} Phase mismatch $\Delta k(\omega_S,\omega_I)$ with fixed idler (solid) or signal (dashed) wavelengths in the 1200-nm to 1800-nm range. The intersection points between the solid and the dashed lines of the same color are where a first-order approximation for the phase mismatch is performed. The black dotted line is formed by all intersections at different wavelengths, giving the phase mismatch, $\Delta k_0$, at the degenerate wavelengths. }
\end{figure}

\subsection{The GVM condition and Gaussian phase-matching profile}
To generate spectrally-uncorrelated biphotons, we desire $f(\omega_S,\omega_I)= \alpha(\omega_S + \omega_I) G(\Delta k(\omega_S,\omega_I))= \beta_S(\omega_S)\beta_I(\omega_I)$. A conventional approach to engineer such a product-state wave function is picking the signal and idler wavelengths, $\omega_{S_0}$ and $\omega_{I_0}$, that satisfy $\partial\Delta k/\partial\omega_S \vert_{\omega_{S_0}}=-\partial\Delta k/\partial\omega_I \vert_{\omega_{I_0}} = \gamma_{\rm GVM}$, known as the GVM condition. Around $\omega_{S_0}$ and $\omega_{I_0}$, the phase mismatch is fully determined by the frequency difference between the signal and idler photons, i.e., $\Delta k\approx\Delta k_0 + \gamma_{\text{GVM}}(\omega_S-\omega_I)$. This leads to a phase-matching function $G(\Delta k)$ that is solely a function of $\omega_S - \omega_I$, viz. $G(\omega_S-\omega_I)$. Under the GVM condition, it is possible to engineer a Gaussian phase-matching function 
\begin{equation}
    G(\omega_S - \omega_I) \approx G_0\exp\left[-\frac{(\omega_S - \omega_I)^2}{\sigma^2_C}\right],
\end{equation}
where $\sigma_C$ is determined by the poling profile $g(z|\boldsymbol{A}_{\rm GVM})$. In conjunction with a Gaussian pump spectral profile 
\begin{equation}
   \alpha(\omega_P) = a_0 \exp \left[-\frac{(\omega_S + \omega_I-\omega_{P_0})^2}{\sigma^2_P}\right] 
\end{equation}
and by choosing $\sigma_P = \sigma_C$, one obtains 
\begin{equation}
\begin{aligned}
&f(\omega_S,\omega_I) = \alpha(\omega_S+\omega_I)G(\omega_S - \omega_I) \\
&\propto \exp\left[-\frac{2(\omega_S-\omega_{S_0})^2}{\sigma_P^2}\right]\exp\left[-\frac{2(\omega_I-\omega_{I_0})^2}{\sigma_P^2}\right],  
\end{aligned}
\end{equation}
i.e., a spectrally-uncorrelated product state of biphotons.

The JSA engineering approach based on GVM and Gaussian phase matching, albeit ingenuous, is limited by the dispersion properties of the nonlinear materials, resulting in only a handful of GVM wavelengths, each associated with a specific nonlinear optical material. For example, the GVM wavelengths are $\sim$1582 nm for KTP, $\sim$830 nm for KDP ($\text{KH}_2\text{PO}_4$), and $\sim$922 nm for ADA ($\text{NH}_4\text{H}_2\text{AsO}_4$) \cite{jin2019theoretical}. Such a restriction impedes the generation of spectrally-uncorrelated biphotons that covers the entire telecommunication band from $\sim$1300 nm to $\sim$1600 nm and precludes interfacing SPDC photons with solid-state quantum emitters in the visible to near-infrared wavelength range, as a means to entangle qubits at a distance. To engineer spectrally-uncorrelated biphoton JSAs over a wide spectral range, let us further understand the limitations of the GVM approach.

\subsection{General phase matching without the GVM condition}
\label{sec:generalPM}
Let us now consider working at non-GVM wavelengths. Let the phase mismatch at the degenerate wavelength be $\Delta  k_0$. The phase mismatch at any wavelengths in the vicinity of the degenerate wavelength can be expressed, in a first-order approximation, as $\Delta  k -\Delta  k_0 \approx \gamma_S\omega_S-\gamma_I\omega_I$, where $\gamma_S \neq \gamma_I$ for a non-GVM case. Without loss of generality, both $\gamma_S$ and $\gamma_I$ are chosen positive. The validity of the first-order approximation is verified in Fig.~\ref{fig:DK}, which shows the linearity of $\Delta  k$ around the degenerate wavelengths at the intersections between the solid and dashed lines. Applying the poling profile $\boldsymbol{A}_{\rm GVM}$ developed at the GVM wavelengths, 
\begin{equation}
\label{eq:PMnonGVM}
\begin{aligned}
    &G(\Delta  k,\boldsymbol{A}_{\rm GVM}) \approx G'_0\exp\left[-\frac{(\Delta  k -\Delta  k_0)^2}{\gamma_{\rm GVM}^2\sigma_C^2}\right]\\
    &\times \left[1+\frac{\Delta  k-\Delta  k_0}{\Delta  k_0}+\mathcal{O}((\Delta  k-\Delta  k_0))^2\right]\\
    &\approx G'_0\exp\left[-\frac{(\gamma_S\omega_S-\gamma_I\omega_I)^2}{\gamma_{\rm GVM}^2\sigma_C^2}\right]\\
    &\times \left[1+\frac{\gamma_S\omega_S-\gamma_I\omega_I}{\Delta  k_0}+\mathcal{O}((\gamma_S\omega_S-\gamma_I\omega_I)^2)\right],
\end{aligned}
\end{equation}
where the second term includes high-order contributions. Since $(\gamma_S\omega_S-\gamma_I\omega_I)/\Delta  k_0$ is small but not negligible, it causes a reduced spectral purity. The spectral purity will, nonetheless, still be high after applying narrowband filters as the higher-order term primarily contribute to areas of large detuning from the central wavelength. It appears that by choosing the pump bandwidth $\sigma_P=\gamma_{\rm GVM}\sigma_C/\sqrt{\gamma_S\gamma_I}$, one obtains, in a non-GVM regime, a product-state JSA $f(\omega_S,\omega_I) \propto \exp\left[-\frac{(\omega_S-\omega_{S_0})^2}{\sigma_S^2}\right]\exp\left[-\frac{(\omega_I-\omega_{I_0})^2}{\sigma_I^2}\right]$, where 
\begin{equation}
    \begin{aligned}
    \sigma_S &= \frac{\sigma_C \gamma_{\rm GVM}}{\sqrt{\gamma_S(\gamma_S+\gamma_I)}}\\
    \sigma_I &= \frac{\sigma_C \gamma_{\rm GVM}}{\sqrt{\gamma_I(\gamma_S+\gamma_I)}}.
    \end{aligned}
\end{equation}
However, a number of caveats about this result need be noted. First, a practical phase-matching function $G(\Delta k,\boldsymbol{A}_{\rm GVM})$ is only approximately Gaussian due to the finite number of periods and the minimum duty cycle $\Lambda_{\rm min}$ that induces discretization errors; and second, the higher-order Hermit-Gaussian terms in Eq.~\ref{eq:PMnonGVM} generate side lobes, which should be accounted for in an effective design. As a result, a 99.5\% filtered spectral purity \cite{dixon2013spectral} at the GVM wavelength and Gaussian phase matching reduces to 97.13\% in the 1550-nm band after applying a 40-nm filter. 

\section{The machine-learning framework}
\label{sec:MLFramework}

\begin{figure}
\centering
\includegraphics[width=8cm]{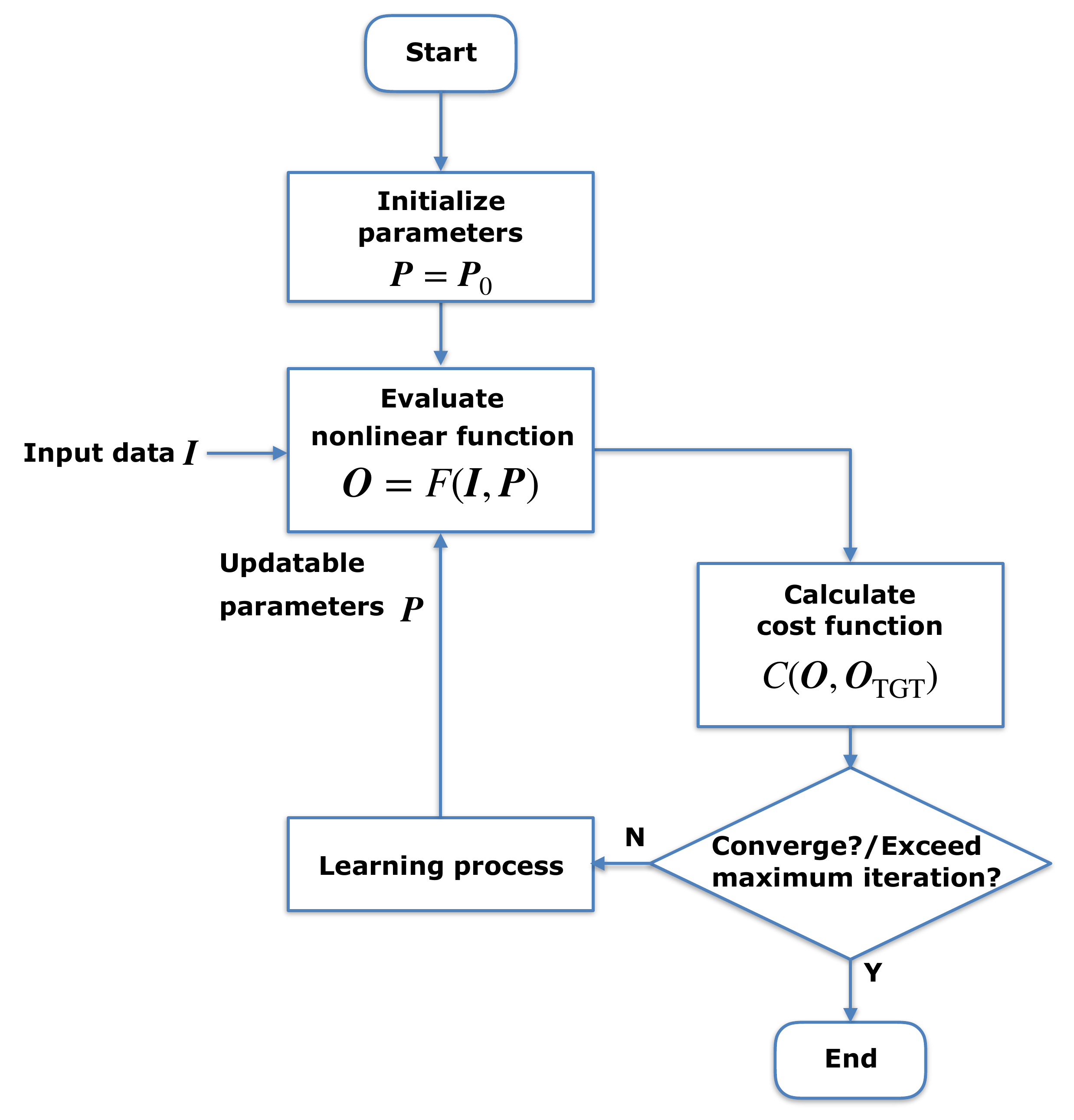}
\caption{A general machine-learning model. $\boldsymbol{I}$: input data; $\boldsymbol{O}$: output data; $\boldsymbol{P}$: updatable parameters; $\boldsymbol{P}_0$: initial parameters; $F(\boldsymbol{I},\boldsymbol{P})$ nonlinear function; $\boldsymbol{O}_{\rm TGT}$: target output data; $C(\boldsymbol{O}, \boldsymbol{O}_{\rm TGT})$: cost function that determines the distance between the present output data and the target output data; Learning process is an algorithm that updates the parameters based on the cost.}
\label{fig:MachineLearning} 
\end{figure}

We introduce a machine-learning framework to cope with the limitations associated with the GVM and Gaussian-phase-matching approach. The machine-learning framework enables the suppression of higher-order terms and compensations for discretization errors, leading to high-spectral-purity biphotons over a wide range of wavelengths. In addition, the machine-learning framework is capable of designing a poling profile that corrects deviations from a perfect Gaussian-spectrum pump.

A general machine-learning framework is comprised of multiple building blocks, as illustrated in Fig.~\ref{fig:MachineLearning}. A nonlinear function $F$ is characterized by a set of updatable parameters $\boldsymbol{P}$. $F$ takes input data $\boldsymbol{I}$ and produces output data $\boldsymbol{O} = F(\boldsymbol{I},\boldsymbol{P})$. The objective of learning is to seek the optimum $\boldsymbol{P}$ so that $\boldsymbol{O}$ converges to the target output data $\boldsymbol{{O}}_{\rm TGT}$ for different $\boldsymbol{I}$. To this end, we define a cost function, $C(\boldsymbol{O}, \boldsymbol{{O}}_{\rm TGT})$, that quantifies the distance between $\boldsymbol{O}$ and $\boldsymbol{{O}}_{\rm TGT}$. $\boldsymbol{P}$ is initialized by a preset of parameters $\boldsymbol{P}_0$ and updated by a learning process supplied with the calculated cost in each iteration.

The most common class of learning processes are based on gradient descent, in which $\boldsymbol{P}$ is updated based on a linear scaling of the negativity of the gradient of the cost function \cite{ruder2016overview,le2011optimization}. The linear scaling is defined as the learning rate. {\em Adam} is an upgraded version of the gradient-descent learning process. {\em Adam} introduces the adaptive momentum method that adjusts the learning rate based on the learning history \cite{kingma2014adam, 1807.06766}. Such a feature differentiates {\em Adam} from a conventional gradient-descent learning process with a fixed learning rate.

To date, several optimization approaches based on {\em forwardpropagation} have been employed to design the poling profile for JSA engineering \cite{dosseva2016shaping, tambasco2016domain, graffitti2017pure}. These approaches rest upon equal poling periods and static strategies to optimize the poling profile. In contrast, gradient descent and {\em Adam} fall into {\em backpropagation} learning processes because the gradient of the cost function is calculated as a partial derivative over elements in $\boldsymbol{P}$. Critically, the backpropagation approach in our machine-learning framework automates the optimization procedure. Moreover, the search time in the parameter space is significantly reduced by virtue of {\em Adam}'s adaptive parameter update strategy.

\begin{figure}[htbp]
\centering
\includegraphics[width=8cm]{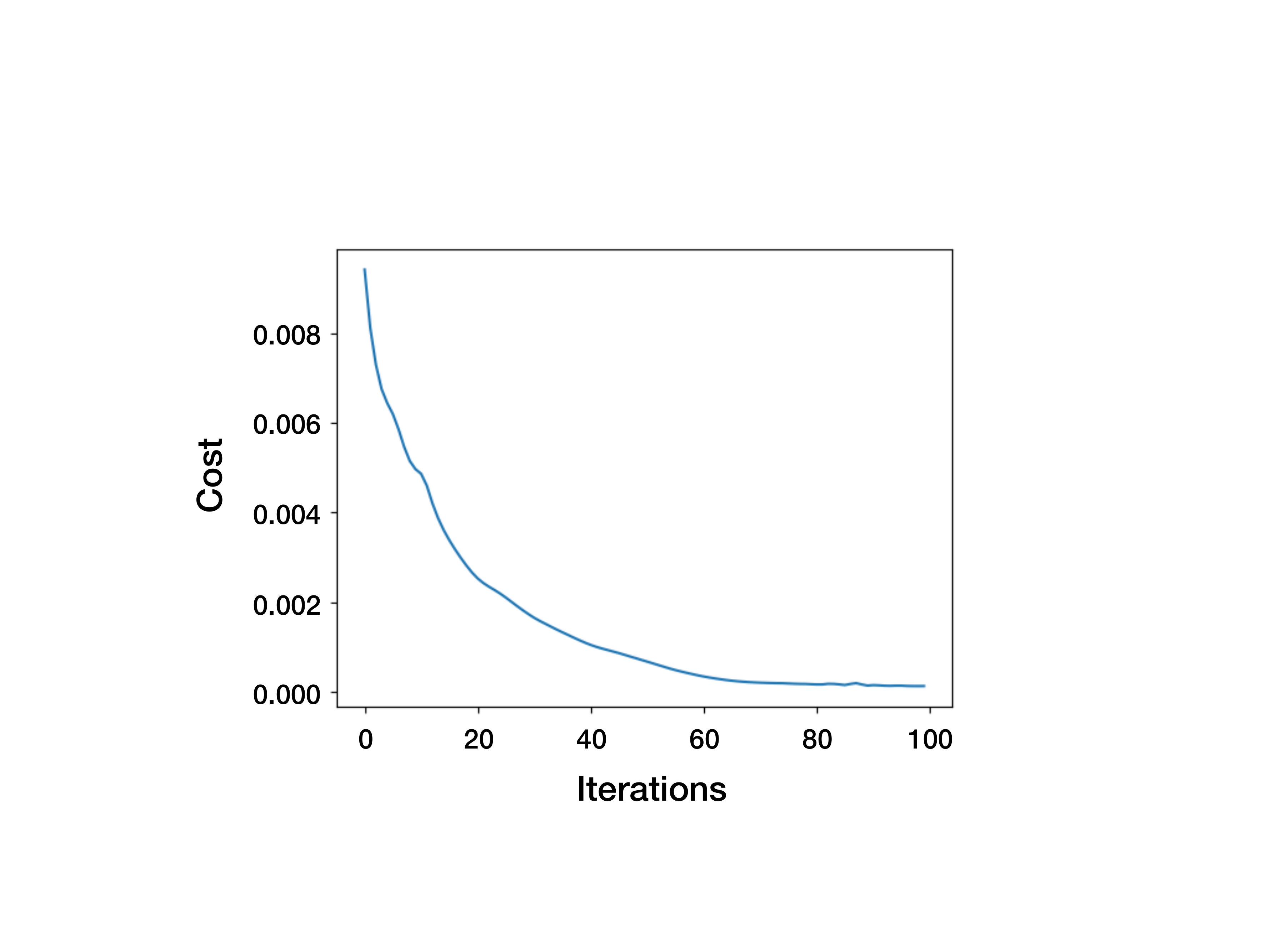}
\caption{\label{fig:Cost} The evolution of the cost during poling optimization for the generation of spectrally-uncorrelated biphotons in the 1550-nm band. The cost approaches $0$ after 100 iterations, indicating near-spectrally-uncorrelated biphotons can be produced with the optimized crystal poling.}
\end{figure}

To utilize the machine-learning framework to engineer the JSAs for spectrally-uncorrelated biphotons, we assume that a pump with an ideal Gaussian spectrum and bandwidth is available at the desired wavelength, with the understanding that deviations from the ideal conditions can be accommodated by the machine-learning framework. Consider the function $H(\omega_S,\boldsymbol{A}) \equiv |G(\Delta k(\omega_S, \omega_{P_0} - \omega_I),\boldsymbol{A})|$, i.e., a slice of the phase-matching function $|G(\Delta k(\omega_S,\omega_I),\boldsymbol{A})|$ along the $\omega_S + \omega_I=\omega_{P_0}$ axis, where $\omega_{P_0}$ is the central frequency for the pump. If $H(\omega_S,\boldsymbol{A})$ is Gaussian, $|G(\Delta k(\omega_S, \omega_I),\boldsymbol{A})|$ will have the form as the first term in Eq.~\ref{eq:PMnonGVM}. Therefore, Our objective is to seek the optimum poling profile $\boldsymbol{A}$ such that $ H(\omega_S,\boldsymbol{A}) \rightarrow H_0 \exp\left[-\frac{(\omega_S-\omega_{T_0})^2}{2\sigma_T^2}\right]$, where $\omega_{T_0}$ and $\sigma_T$ is the central frequency and the standard deviation for this Gaussian target. The purity of the biphotons ties to distance between $ H(\omega_S,\boldsymbol{A})$ and an ideal Gaussian form. In the machine-learning framework, the input data $\boldsymbol{I}$ consist of an array of sampled signal frequencies: $\boldsymbol{\omega_S} = \{\omega_{S_1},\omega_{S_2},...,\omega_{S_k}\}$. The updatable parameters $\boldsymbol{P}$ include the duty cycle array $\boldsymbol{A}$, the target central frequency $\omega_{T_0}$, and the target bandwidth $\sigma_{T}$. With the updatable parameters, the nonlinear function returns the output data $\boldsymbol{O} = F(\boldsymbol{I},\boldsymbol{P}) = H(\boldsymbol{\omega_S},\boldsymbol{A})$, which aims to approach the target output data
\begin{equation}
    \boldsymbol{O}_{\rm TGT} = H_0 \exp\left[-\frac{(\boldsymbol{\omega_S}-\omega_{T_0})^2}{2\sigma_T^2}\right]
\end{equation}
The cost is then defined as the distance between the output data and an ideal Gaussian function as
\begin{equation}
\begin{aligned}
    C(\boldsymbol{O},\boldsymbol{O}_{\rm TGT}) &= \sum_{l = 1}^k \bigg\{H(\omega_{S_l},\boldsymbol{A})\\ 
    &- H_0 \exp\left[-\frac{(\omega_{S_l}-\omega_{T_0})^2}{2\sigma_T^2}\right]\bigg\}^2
\end{aligned}
\end{equation}

\begin{figure}[t]
\centering
\includegraphics[width=8cm]{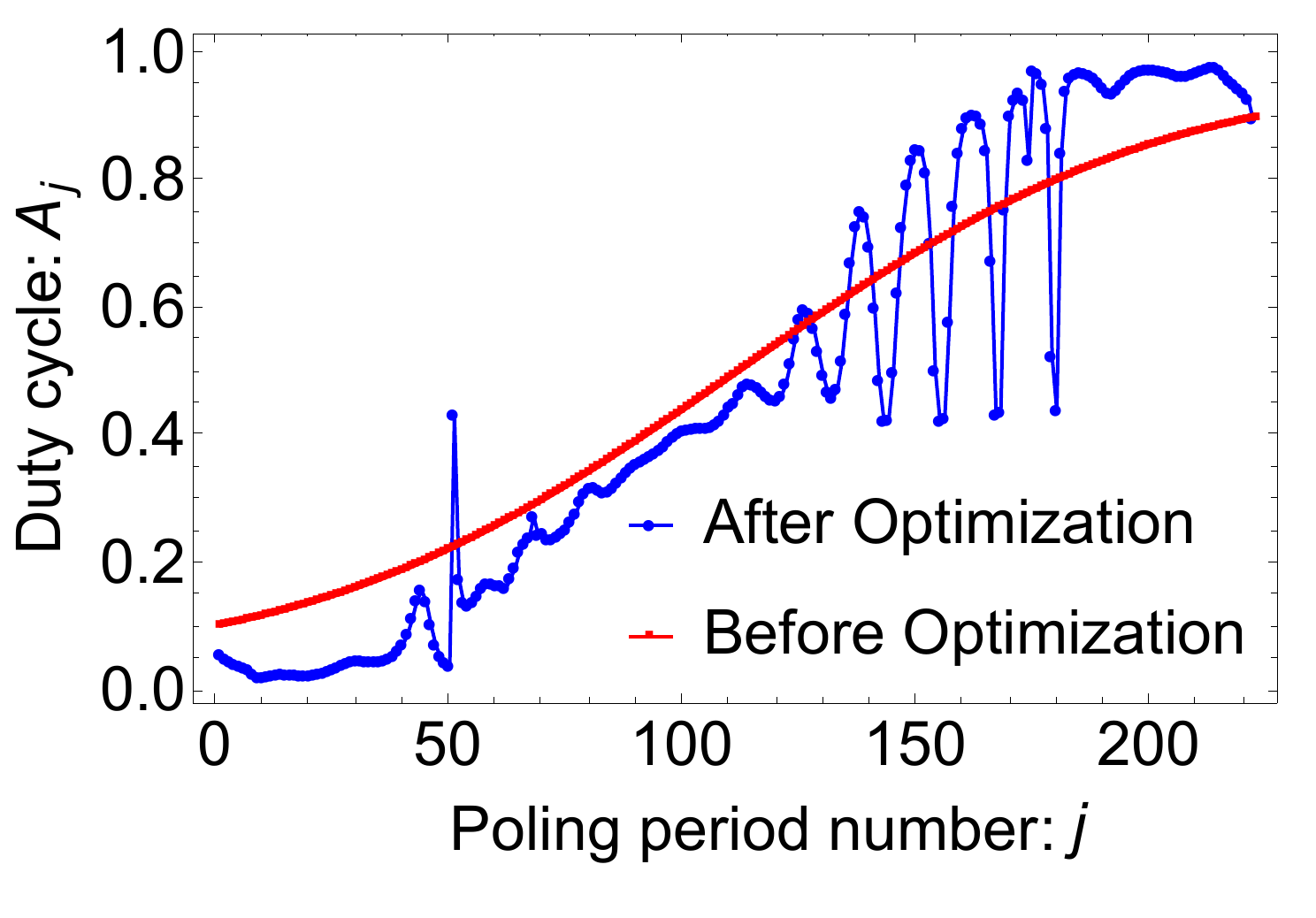}
\caption{\label{fig:PP} Poling profiles for the generation of spectrally-uncorrelated biphotons in the 1550-nm band. Red curve: the initial poling profile prior to applying machine learning, obtained from the GVM condition and Gaussian phase matching \cite{dixon2013spectral}; Blue dots: the optimized poling profile obtained by the machine-learning framework.}
\end{figure}

\section{Poling-design recipe}
\begin{figure*}[t]
\centering
\includegraphics[width=14cm]{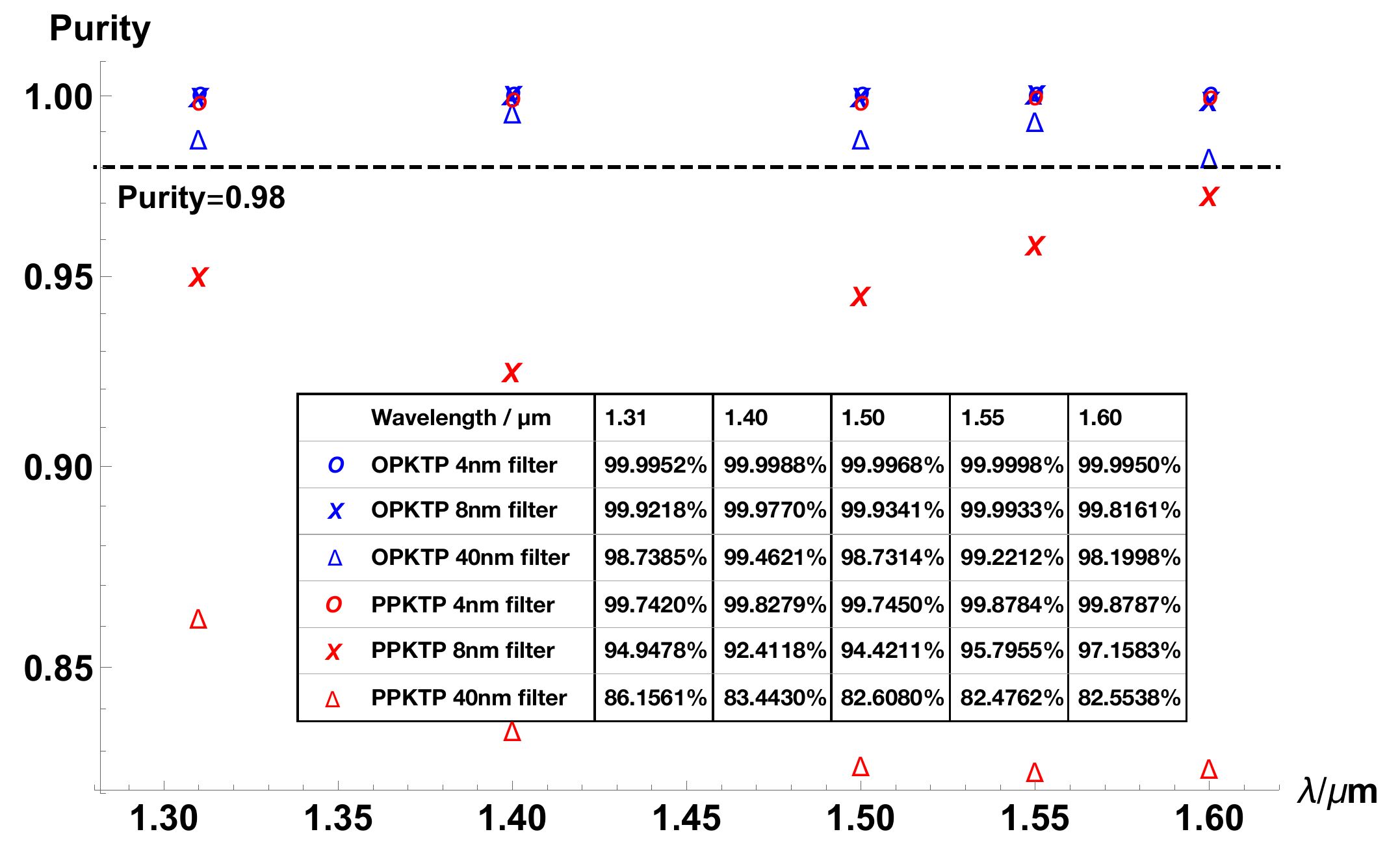}
\caption{\label{fig:Purity} Simulated purities at different wavelengths after applying filters of different bandwidths. Circles: purities with our machine-learning-based poling design; Crosses: purities with periodic poling. For PPKTP, a larger wavelength yields larger bandwidth for the biphotons. A 8-nm filter then eliminates the first side lobe to increase the purity goes. Inset: a table that summarizes the purities for both PPKTP and OPKTP.}
\end{figure*}

\begin{figure*}[t]
\centering
\includegraphics[width=14cm]{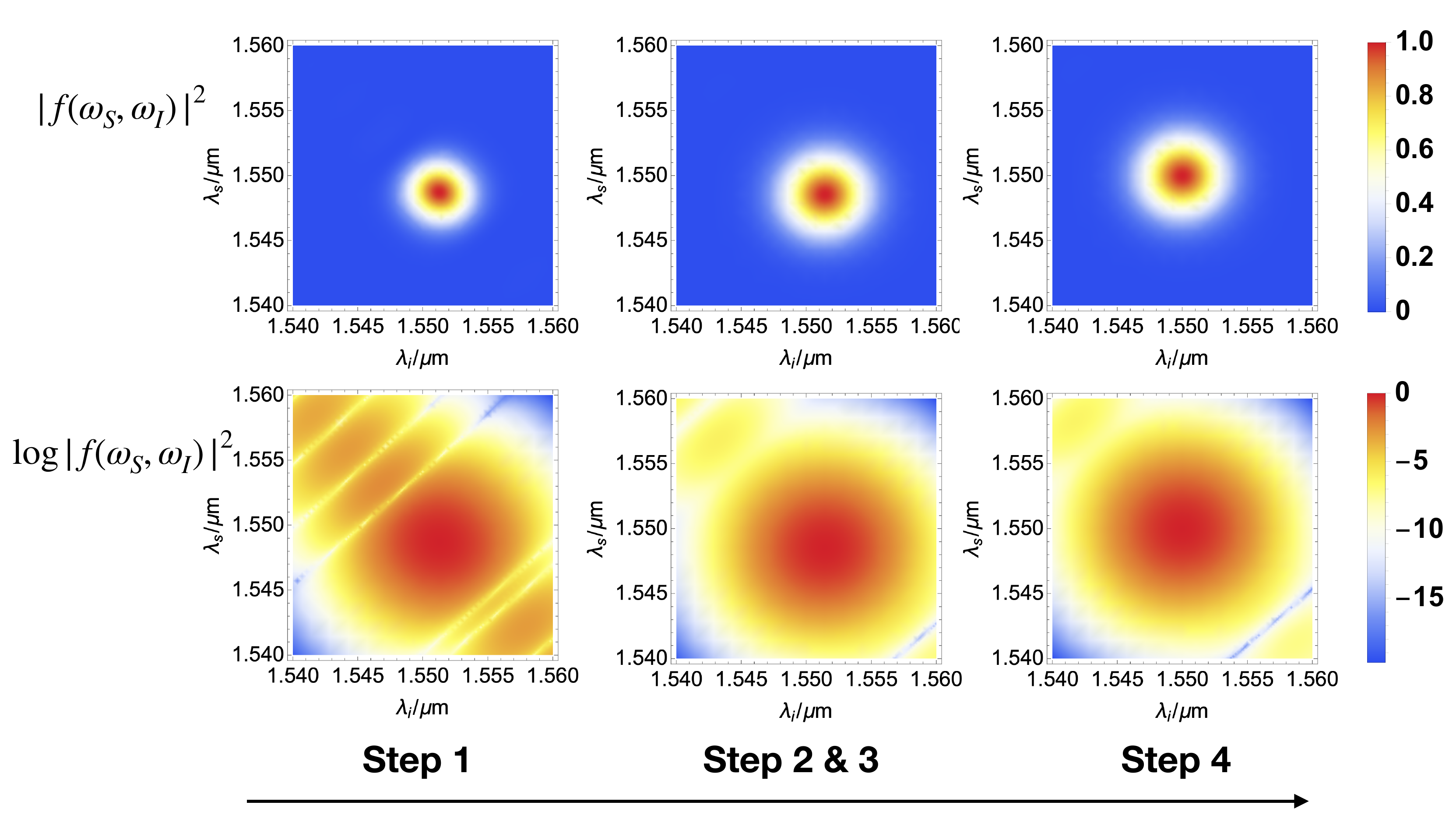}
\caption{\label{fig:Steps} Plot for $|f(\omega_S,\omega_I)|^2$, and $\log|f(\omega_S,\omega_I)|^2$ after each design step. Step 1 takes a prior poling design, which, without any optimization, suffers from side lobes that degrade the purity. After Step 2\&3, the side lobes are suppressed by the machine-learning framework while leaving the biphotons non-degenerate. After Step 4, the biphotons become degenerate at 1550 nm. The logarithm scale amplifies the visibility of the side lobes.}
\end{figure*}

We customize the poling for an $L=$ 10-mm KTP crystal comprised of $\sim$200 poling periods. Such a size of the parameter space would be a challenge for any analytic attempt to devise the optimum poling profile. The machine-learning framework, by contrast, optimizes the duty cycle array $\boldsymbol{A}$ by gradually minimizing the cost function over each iteration. In doing so, the optimization does not rely on any approximation, nor does it place any requirement on the pump profile. It can thus cope with any target output data under any pump profile. In what follows, we describe the four main steps of our poling-design recipe.

\textbf{Step 1: Initialization } Set the target wavelength and the bandwidth of interest. Set the poling period $\Lambda=2\pi/\Delta k_0$ at the target wavelength. Set the minimum poling length $A_{\rm min}$ allowed by the fabrication processes. As a reference, $A_{\rm min}=0$ is an upper bound for the performance. Initialize $\boldsymbol{A}=\boldsymbol{A}_{\rm GVM}$ and choose the maximum number of machine-learning iterations.\\

\textbf{Step 2: Cost function calculation } Obtain and normalize $\boldsymbol{O} = H(\boldsymbol{\omega_S},\boldsymbol{A})$. Calculate the cost function $C(\boldsymbol{O},\boldsymbol{O}_{\rm TGT})$.\\

\textbf{Step 3a: Pump optimization } Use {\em Adam} with learning rate $r_a$ to update $\omega_{T_0}$ and $\sigma_T$ and reduce the cost function.\\

\textbf{Step 3b: Poling optimization } Use {\em Adam} with learning rate $r_b$ to update $\boldsymbol{A}$ and reduce the cost function. Repeat from \textbf{Step 2} until the maximum number of iterations is reached or the cost function converges to the minimum.\\

\textbf{Step 4: Poling period adjustment } Fine tune $\Lambda$ without modifying $\boldsymbol{A}$ to eliminate the non-degeneracy between the signal and idler photons.

\section{Simulation results}
The machine-learning framework for the poling design is realized in Python with the TensorFlow library \cite{code}. The material dispersion profile of KTP is derived by the Sellmeier equation reported in Ref.~\cite{kato2002sellmeier}. The employed learning rates are $r_a=0.005$ and $r_b=0.015$.

Working in the 1550-nm band, the cost converges after 100 machine-learning iterations, shown in Fig.~\ref{fig:Cost}. Fig.~\ref{fig:PP} displays the initial poling profile and the optimized poling profile obtained by the machine-learning framework. The periodic peaks in the machine-learned poling profile may be responsible for the compensation of discretization errors. After applying filters with various bandwidths, the spectral purity of the produced biphotons is calculated by the coefficients of Schmidt decomposition\cite{dixon2013spectral,ansari2018tailoring}. We apply our recipe to generating spectrally-uncorrelated biphotons in the 1310-nm, 1400-nm, 1500-nm, and 1600-nm bands. The spectral purities with filters with various bandwidth are derived for both optimized-poling KTP (OPKTP) and PPKTP, as depicted in Fig.~\ref{fig:Purity} and summarized in the inset. The machine-learning-based poling design clearly improves the purity of biphotons.

To further illustrate the poling optimization procedure, Fig.~\ref{fig:Steps} shows the joint spectral intensity, defined as $|f(\omega_S,\omega_I)^2|$, in the 1550-nm band after each step. One observes that side lobes are suppressed after Step 2 \& 3, and Step 4 eliminates the non-degeneracy between the biphotons.

Table.~\ref{tab:Table} collects key metrics reported in this and prior works of poling design for spectrally-uncorrelated photon generation. Since different works employ filters of different bandwidths and set different resolutions for JSAs in calculating the purity, the highest purity numbers do not accurately reflect the performance for different schemes. Notably, our machine-learning framework is able to compensate higher-order terms by direct calculating the cost function from the sampled biphoton JSA, whereas prior works are all based on first-order approximations, as discussed in Sec.~\ref{sec:generalPM}. Because our machine-learning framework relies on neither a first-order approximation nor the GVM condition, high spectral purity can be achieved over the entire telecommunication band from 1300 nm to 1600 nm. We should note that the biphoton JSA centered at non-GVM wavelengths, e.g., a central wavelength at 1310 nm, is elliptical since $\sigma_S\neq\sigma_I$. The unequal bandwidth of the signal and idler photons however does not prevent signal or idler photons from independent sources to interfere, as a building block for entanglement swapping and quantum teleportation.

\begin{table*}[htbp]
\centering
\resizebox{\textwidth}{!}{%
\begin{tabular}{|c|c|c|c|c|c|c|c|}
\hline
{\bf Reference} & {\bf Poling strategy} & {\bf Optimization method} & \begin{tabular}[c]{@{}c@{}}\\ {\bf Central} \\ {\bf wavelength} \\ \\ \end{tabular} & \begin{tabular}[c]{@{}c@{}} {\bf Domain} \\ {\bf number}\end{tabular} & \begin{tabular}[c]{@{}c@{}}{\bf Crystal} \\ {\bf length}\end{tabular} & \begin{tabular}[c]{@{}c@{}}{\bf Filter type}\\ \& {\bf bandwidth} \end{tabular} & \begin{tabular}[c]{@{}c@{}} {\bf Achieved} \\ {\bf highest purity}\end{tabular} \\ \hline
M. Bra{\'n}czyk \textit{et al.} \cite{branczyk2011engineered} & \begin{tabular}[c]{@{}c@{}}\\ Customized \\ poling order\\\end{tabular} & Analytic design & 1576 nm & $\sim$900 & 24.2 mm & N/A & 99.0\% \\
P. B. Dixon \textit{et al.} \cite{dixon2013spectral} & \begin{tabular}[c]{@{}c@{}}\\ Customized \\ duty cycle\\\end{tabular} & Analytic design & 1582 nm & $\sim$520 & 12 mm & \begin{tabular}[c]{@{}c@{}}Gaussian \\ 8.5 nm\end{tabular} & 99.5\% \\
J. Tambasco \textit{et al.} \cite{tambasco2016domain} & \begin{tabular}[c]{@{}c@{}}\\ Customized \\ orientation\\\end{tabular} & Binary coordinate descent & 1550 nm & $\sim$532 & 12 mm & \begin{tabular}[c]{@{}c@{}} Rectangular \\  $\sim$16 nm \cite{Tnote}\end{tabular} & 99.6\% \\
A. Dosseva \textit{et al.} \cite{dosseva2016shaping} & \begin{tabular}[c]{@{}c@{}}\\Customized \\ orientation\\\end{tabular} & Binary simulated annealing & 1582 nm & 1300 & $\sim$14.1 mm & \begin{tabular}[c]{@{}c@{}} Rectangular \\ $\sim$10 nm \cite{Tnote}\end{tabular} & 99.9\% \\
F.  Graffitti \textit{et al.} \cite{graffitti2017pure} & \begin{tabular}[c]{@{}c@{}}\\Customized orientation \\ + tuning periods\\\end{tabular} & \begin{tabular}[c]{@{}c@{}}Binary coordinate descent \\ \& simulated annealing\end{tabular} & 1582 nm & $\sim$870 & 2 mm & N/A & 99.0\% \\
F.  Graffitti \textit{et al.} \cite{graffitti2017pure} & \begin{tabular}[c]{@{}c@{}}\\Sub-coherence\\ engineering\\\end{tabular} & Binary coordinate descent & 1582 nm & $\sim$870 & 2 mm & N/A & 99.4\% \\
This work & \begin{tabular}[c]{@{}c@{}}\\Customized duty cycle\\ + tuning periods\\ \\\end{tabular} & {\em Adam} & \begin{tabular}[c]{@{}c@{}}1300nm \\     -1600 nm\end{tabular} & \begin{tabular}[c]{@{}c@{}}$\sim$400 \\     -$\sim$500\end{tabular} & 10 mm & \begin{tabular}[c]{@{}c@{}}Rectangular\\ 8 nm\end{tabular} & \textgreater 99.8\% \\ \hline
\end{tabular}%
}
\caption{Performance-metric comparison for prior works and the present work.}
\label{tab:Table}
\end{table*}

\section{Discussion}
The machine-learning framework represents a general optimization strategy particularly suitable for complex problems that have clear objectives but cannot be tackled by approximation methods due to the sensitivity to small variations in their solutions. This kind of problems typically involves a large parameter space that is hard to solve by conventional optimization methods. Intriguingly, the solutions sought by the machine-learning framework may in turn offer new insights for the complex problems in hand.

The machine-learning framework utilizes {\em Adam} as the learning process. Since {\em Adam} converges at a zero-gradient minimum point, the poling profile is robust against small fabrication errors. {\em Adam} has shown great performance in many non-convex optimization problems. To fully unleash the potential of {\em Adam}, it sometimes requires careful choice of the learning rate. Since one cannot ensure the convergence of the machine-learning algorithm to the global optimum, it is recommended to set initial parameters based on a good existing design and subsequently leverage machine learning to achieve substantial improvement. In this work, we set a known Gaussian poling profile obtained at the GVM wavelength as the initial parameters for the machine-learning framework. In the learning process, a finer sampling resolution will slightly improve the performance at the cost of requiring more computational resources. Apart from KTP, our machine-learning-based design recipe is applicable to nonlinear crystals, as long as a first-order approximation dominates higher-order terms in the phase-mismatch function at the working wavelength.

The machine-learning framework can also be used to seek poling profiles for other forms of biphoton JSAs by simply modifying the target JSA in the cost function. For example, non-degenerate biphoton states and non-Gaussian states can be engineered by our approach, as a means for entangling solid-state qubits at a distance \cite{sangouard2011quantum,halder2007entangling,lu2019chip}. To engineer an arbitrary biphoton JSA, a learning process solely based on sampling the signal frequencies becomes insufficient. In such a general situation, one should sample both the signal's and the idler's frequencies, in $\boldsymbol{\omega_S}$ and $\boldsymbol{\omega_I}$, and feed to the nonlinear function to obtain $H(\boldsymbol{\omega_S},\boldsymbol{\omega_I}|\boldsymbol{A})$ as the output. The cost function should also be modified accordingly. There is, however, no fundamental constraints that prevent the machine-learning framework from engineering an arbitrary biphoton JSA, but given a finite number of poling periods, the machine-learning framework may only approach a target biphoton JSA down to a certain precision \cite{hornik1989multilayer}. 

\section{Conclusions}
We have developed a machine-learning framework to solve the problem of the generation of indistinguishable biphotons in the telecommunication band over 1300-nm to 1600-nm. Our approach leads to a spectral purity in excess of 99.99\% for biphotons in the 1550-nm band after applying a 8-nm filter. This work demonstrates machine learning's potential of advancing quantum information science. We hope that this work will spur the pursuits of other machine-learning-enhanced quantum communication, sensing, and information processing applications.

\section*{Acknowledgments}
This work is funded by National Science Foundation Major Research Instrumentation Program Award\# 1828132 with matching funds provided by the University of Arizona. CC acknowledges the Nicolaas Bloembergen Graduate Student Scholarship. ZZ is grateful for support from the University of Arizona. SG acknowledges the Office of Naval Research program Communications and Networking with Quantum Operationally-Secure Technology for Maritime Deployment (CONQUEST), awarded under Raytheon BBN Technologies prime contract number N00014-16-C2069, and a subcontract to University of Arizona. NP acknowledges support from Regents Innovation Funds. QZ is supported by the University of Arizona.
\bibliography{Ref}

\end{document}